\documentclass[runningheads, 10pt]{llncs}

\usepackage[most]{tcolorbox}
\usepackage{xcolor}

\usepackage{graphicx}

\usepackage{booktabs}
\usepackage{tabularx}
\usepackage{multirow}
\usepackage{enumitem}
\setlist[itemize]{noitemsep, topsep=0pt}
\usepackage{amstext}  
\usepackage{xspace}
\usepackage{amssymb}
\usepackage[overload]{empheq}

\usepackage{caption}
\usepackage{graphbox}

\usepackage{subcaption}

\usepackage{adjustbox}
\usepackage[linesnumbered,algo2e,ruled]{algorithm2e}
\usepackage{multirow}
\usepackage{tikz}

\usepackage{pifont}
\usepackage{svg}
\usepackage{soul}

\usepackage{amsmath}
\usepackage{amsfonts, amssymb}
\usepackage{setspace}
\usepackage{colortbl}
\usepackage{cellspace}

\usepackage[colorlinks=true,linkcolor=blue,breaklinks=True,citecolor=brown,urlcolor=blue]{hyperref}

\renewcommand{\textsc}[2]{{\fontfamily{cmr}\selectfont {#1}{\footnotesize {#2}}}}

\newcommand{\textbox}[1]{
    \noindent\fbox{%
        \parbox{0.97\columnwidth}{%
            {#1}
        }%
    }
}

\newtcolorbox{cooltextbox}[1][]{%
    colback=black!5,
    colframe=black!5,
    notitle,
    sharp corners,
    borderline west={0pt}{0pt}{red!80!black},
    enhanced,
    breakable,
    left=0pt,
    right=0pt,
    top=0pt,
    bottom=0pt
    }

\newcommand\prodigy[0]{{\fontfamily{pcr}\selectfont {\small PRODIGY}}}
\newcommand\scprodigy[0]{{\fontfamily{pcr}\selectfont {\scriptsize PRODIGY}}}

\begin{document}
\title{LLM4PM: A case study on using Large Language Models for Process Modeling in Enterprise Organizations}

\titlerunning{A case study on LLM for Process Modeling in Enterprise Organizations}

\author{Clara Ziche
\and Giovanni Apruzzese
}

\institute{University of Liechtenstein}

\maketitle 

\begin{abstract}
We investigate the potential of using Large Language Models (LLM) to support process model creation in organizational contexts. Specifically, we carry out a case study wherein we develop and test an LLM-based chatbot, \prodigy\ (PROcess moDellIng Guidance for You), in a multinational company, the Hilti Group. We are particularly interested in understanding how LLM can aid (human) modellers in creating process flow diagrams. To this purpose, we first conduct a preliminary user study (n=10) with professional process modellers from Hilti, inquiring for various pain-points they encounter in their daily routines. Then, we use their responses to design and implement \prodigy. Finally, we evaluate \prodigy\ by letting our user study's participants use \prodigy, and then ask for their opinion on the pros and cons of \prodigy. We coalesce our results in actionable takeaways. Through our research, we showcase the first practical application of LLM for process modelling in the real world, shedding light on how industries can leverage LLM to enhance their Business Process Management activities.
\end{abstract}

\section{Introduction}
\label{sec:introduction}

\noindent
Organizations perform business processes to deliver value-adding outcomes to their customers. Hence, Business Process Management (BPM) capabilities, such as \textit{process modeling}, are a pivotal task in modern enterprises~\cite{dumas2018fundamentals}. However, despite decades of efforts~\cite{herbst1999inductive}, process modeling still remains a costly activity due to, e.g., the difficulty of providing clear, up-to-date and easy-to-retrieve \textit{documentation}~\cite{dumas2018fundamentals} to those tasked to carry out such activities---the process modelers.

Inspired by recent developments in artificial intelligence (AI), such as large language models (LLM), researchers have proposed various techniques that can facilitate BPM-related tasks (e.g.,~\cite{grohs2023large}). Indeed, LLM can elaborate large collections of documents. Hence, by receiving an input from a given user, LLM can quickly produce an output that {\small \textit{(i)}}~accounts for existing documentation, while simultaneously {\small \textit{(ii)}}~answering the request of the user---i.e., a human. 
Yet, we found no evidence of \textit{practical} applications of LLM for BPM in real contexts and, in particular, for process modeling. 
Hence, there is a need to investigate the effectiveness of such automation in industry~\cite{plattfaut2022critical_pc}. Here, we tackle this challenge and showcase how a large enterprise, \textbf{Hilti}, can benefit from a LLM-powered chatbot---which we developed ad-hoc for Hilti---for BPM.

\begin{figure}[!htbp]
    \vspace{-0.6cm}
    \centering
    \includegraphics[width=0.95\columnwidth]{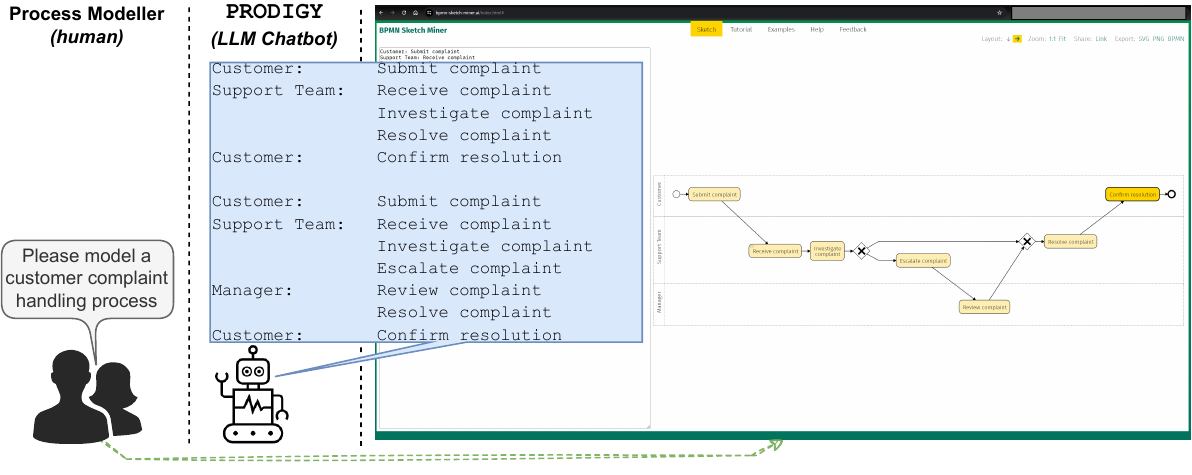}
    \vspace{-0.4cm}
    \caption{\textbf{An exemplary usage of \scprodigy} -- Our LLM-powered chatbot can fulfill various BPM-related tasks. Its most appreciated functionality is generating an output for a complete process model (if copy-pasted into the open-source tool BPMN Sketch Miner~\cite{ivanchikj2022live})}
    \label{fig:example}
    \vspace{-8mm}
\end{figure}

\textbf{\textsc{C}{ONTRIBUTIONS}.} We present a (the first) real-world case study showcasing the application of LLM for Process Modeling in operational contexts. Specifically, to advance the state of the art on BPM, we:
\begin{itemize}[label={\scriptsize $\bullet$},noitemsep,leftmargin=0.3cm]
    \item describe the \textit{problems faced by the considered organization}, Hilti, providing evidence of the necessities of modern enterprises (§\ref{sec:problem});
    \item carry out a \textit{requirement analysis by conducting interviews} with Hilti's employees, shedding light on the pain-points of professional process modelers (§\ref{ssec:artifact_definition});
    \item use our interviews as a scaffold to \textit{develop an original LLM-based chatbot}, \prodigy\ (Fig.~\ref{fig:example}), designed to support Hilti's process modelers (§\ref{ssec:artifact_implementation});
    \item evaluate the ability of \prodigy\ to \textit{generate practical value} by {\small \textit{(i)}}~having Hilti's employees use \prodigy\ and {\small \textit{(ii)}}~collecting and analysing their feedback (§\ref{ssec:artifact_evaluation}).
\end{itemize}
Our results (§\ref{sec:findings}) show that \prodigy\ is generally well-received, and identify room for improvement. We also derive lessons learned that future work can use to drive practical deployment of LLM-based technologies~(§\ref{sec:significance}).

\vspace{1mm}
{\setstretch{0.8}\textbox{{\small \textbf{Industrial Secret and Ethics:} In this paper, we describe some elements pertaining to the internal processes of Hilti: we have been granted permission to share such information. Furthermore, we carry out our user studies ethically: our institutions are aware of this research, our participants have been informed of their rights and of the purpose of their contributions, and we have their consent to post them.}}}

\section{Organizational Context and Problem Statement}
\label{sec:problem}

\vspace{-2mm}

\noindent
Our case organization, Hilti Group, is a multinational company that was founded in 1941 in Schaan, Liechtenstein. It is a world market leader in fastening and demolition technology for construction professionals and provides tools, technologies, software and services to the global construction industry. In 2023, Hilti’s workforce consists of about 33.000 employees in more than 120 countries, making it a highly diverse, distributed organization that operates in complex and competitive markets all over the world. The size, complexity and business model of Hilti make it an ideal use case for testing the capabilities of LLM for process modeling: BPM is essential to ensure cooperation and consistent outcomes within Hilti's ecosystem; furthermore, it is crucial for Hilti to optimize customer-facing processes. Hence, a smooth process modelling is pivotal for Hilti.

\textbf{Challenge.} Hilti has an extensive and heterogeneous documentation landscape which adds to the intrinsically complex nature of process modeling. Hilti's employees spend abundant time searching through such documentation (our interviews revealed an average of $\sim$40 minutes of search \textit{before} modelling a process). Hence, to improve the productivity of their process modellers, and to actively explore innovative technologies, Hilti is interested in novel solutions that facilitate the routines of their employees.

\textbf{Technological Gap.} LLM-based solutions are common (in 2024). However, existing techniques cannot be applied to Hilti's use case. This is because of the \textit{confidential nature} of Hilti's documents: publicly available models (e.g., ChatGPT) should not be able to access Hilti's data to shape their responses; furthermore, even \textit{interacting} with certain LLM (or their APIs) from within Hilti's networks triggers warnings, preventing a reliable usage of these solutions---which are leveraged also by renown prior work, such as \cite{fill2023conceptual,grohs2023large,klievtsova2023conversational,kourani2024process,bellan2022extracting,sola2023activity}. 

\vspace{-2mm}

\begin{cooltextbox}
\textbf{\textsc{O}{UR} \textsc{G}{OAL}.} We seek to design, develop and evaluate an LLM-based solution that facilitates the job of Hilti's process modellers. The development of such a solution should be driven by Hilti's distinctive organizational's context---including its employee's viewpoint, and its existing documentation.
\end{cooltextbox}

\section{Research and Methods}
\label{sec:method}

\vspace{-2mm}

\begin{figure}[!htbp]
    \vspace{-0.6cm}
    \centering
    \includegraphics[width=0.95\columnwidth]{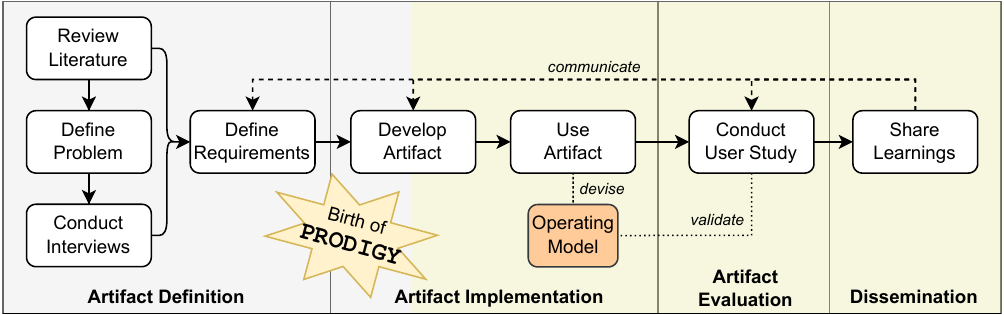}
    \vspace{-0.2cm}
    \caption{\textbf{Method}. We rely on design science research to design, develop, and deploy our artifact. During the implementation of \scprodigy, we also devise an ``operating model'' (which we validate in the evaluation of \scprodigy) through which we explain how \scprodigy\ should be used in real organizations.}
    \label{fig:method}
    \vspace{-0.9cm}
\end{figure}

\noindent
Inspired by Peffers et al.~\cite{peffers2007design}, we followed a Design Science Research (DSR) process consisting of four phases depicted in Fig.~\ref{fig:method}.\footnote{\textbf{Background:} DSR is a methodology that focuses on creating and evaluating artifacts to solve complex problems. Such procedure is rooted on the coming together of people, organizations and technology, with the ultimate intention of ``extending the boundaries of human and organizational capabilities''~\cite{hevner2004design}.} DSR is appropriate given our goal of examining LLMs for process modelling in organizations, as DSR emphasizes the creation of innovative solutions (in our case, \prodigy) while also considering the context in which these solutions will be applied.

\subsection{Artifact Definition}
\label{ssec:artifact_definition}

\vspace{-2mm}

\noindent
As a preliminary step, we carried out a systematic literature review~\cite{vom2015standing} which we used as a foundation to investigate the state of the art and define the scope of our project (see §\ref{sec:problem}). Then, we carried out structured interviews~\cite{franch2023state} meant to identify pain-points and desiderata by professional process modelers\footnote{This is in stark contrast with a closely related work that does not carry out any user study~\cite{kourani2024process}, thereby preventing to fully capture the organizational context.} working for Hilti. We found an agreement with 10 employees, summarised in Table~\ref{tab:population}. The complete questionnaire is provided in our repository~\cite{repository}. Among the most relevant questions, we ask: ``what \textit{challenges} do you experience when modelling processes?'',  ``how helpful is \textit{existing documentation} when you model processes?'' and ``what would you like to see in a new AI artifact that supports process modelling?''; we also provide a list of functionalities for the AI artifact and ask to rate them on a 1--5 scale, as well as potential concerns. Finally, we inquire about the time spent looking for, and reviewing, existing documentation.

\vspace{-9mm}

\begin{table}[!htbp]
    \centering
    \caption{\textbf{Overview of process modellers.} Our participants pertain to various geographical locations of Hilti, and have diverse backgrounds. \textit{[Demographics]} Each participant has 5--25 years of experience in BPM, and they are within 26--60 years of age. Seven hold a MSc. degree. The male:female ratio is 6:4. They all have ``above average'' or ``advanced'' computer knowledge, and all have a basic understanding of LLM. Five perform process modelling activities at least weekly.}
    \label{tab:population}
    \resizebox{0.94\columnwidth}{!}{
    \begin{tabular}{|c|l|l|l|}
    \hline
        \textbf{\#} & \textbf{Job Title} & \textbf{Functional Area} & \textbf{Location}\\ \hline
        \textbf{1} & Business Process Excellence Manager & Corporate & Schaan, FL \\ \hline
        \textbf{2} & Global Process Manager & Communications & Schaan, FL \\ \hline
        \textbf{3}  & Business Process Excellence Manager & Quality Management & Schaan, FL \\ \hline
        \textbf{4} & Business Process Excellence Senior Manager & Corporate & Schaan, FL \\ \hline
        \textbf{5} & Business Process Excellence Expert & Corporate & Schaan, FL \\ \hline
        \textbf{6} & Regional Process Manager & Customer Service & Plano, US\\ \hline
        \textbf{7} & Regional Process Manager & Logistics & Kaufering, GER \\ \hline
        \textbf{8} & Business Process Excellence Lead & Corporate & Schaan, FL\\ \hline
        \textbf{9} & Global Process Manager & Repair & Schaan, FL\\ \hline
        \textbf{10} & Global Process Manager & Repair & Schaan, FL\\ \hline
    \end{tabular}}
\end{table}
\vspace{-0.9cm}

\subsection{Artifact Implementation}
\label{ssec:artifact_implementation}

\vspace{-2mm}

\noindent
We use the results of our interviews alongside those of our investigation of the state of the art to define the requirements of our technical artifact, i.e., the LLM-based chatbot \prodigy. 
To develop \prodigy, we rely on \href{https://web.archive.org/web/20240618004017/https://botpress.com/}{Botpress}, a platform to build custom AI chatbots powered by GPT-based LLMs; for our prototype version of \prodigy, we used GPT-3.5 Turbo, which we found provided satisfactory performance while also requiring less resources to generate an output. 

A crucial aspect of \prodigy\ is its reliance on the \href{https://web.archive.org/web/20240701162110/https://www.bpmn-sketch-miner.ai/index.html}{BPMN Sketch Miner} tool~\cite{ivanchikj2022live}. The syntax for this tool is entirely text-based, human-readable and light in terms of token consumption, making it appropriate for our case study. Therefore, we use few-shot prompting to teach \prodigy\ to provide an output that matches the format expected by BPMN Sketch Miner. This output serves as the input for the model generation and transformation pipeline of BPMN Sketch Miner~\cite{ivanchikj2020text}. Such a design choice enables users of \prodigy\ to directly paste the AI outputs into the online tool and get their model visualized (see Fig.~\ref{fig:example}).

Furthermore, we have leveraged retrieval-augmented generation (RAG)~\cite{lewis2020retrieval} to embed Hilti's documentation into \prodigy. Such documentation included: process descriptions from Hilti's internal documentation repository (anonymised); and information about Hilti's process management, and how to model processes at Hilti (taken verbatim from the learning platform for Hilti's process modellers). These procedures enabled us to instill some knowledge about Hilti's processes in \prodigy---a functionality that was heavily endorsed by our interviewees.

\vspace{1mm}
\textbox{{\small \textbf{Demonstration:} We recorded a \href{https://github.com/Nouronihar/BPM24_LLM4PM/blob/main/PRODIGY_demo.mp4}{video}~\cite{repository} showing the functionalities of \prodigy.}}

\subsection{Artifact Evaluation}
\label{ssec:artifact_evaluation}

\vspace{-2mm}

\noindent
We conducted a user study with our artifact and process modellers. Our aim was to answer evaluative questions on the quality of \prodigy\ for Hilti.

First, the process modellers tested all functionalities of PRODIGY by creating custom prompts, with the intention of simulating their routine tasks. Their inputs and the corresponding AI-generated outputs are \href{https://github.com/Nouronihar/BPM24_LLM4PM/blob/main/annex.pdf}{fully observable} in our repository~\cite{repository}. Then, we carried out semi-structured interviews during which the participants answered 28 questions. Among these, we ask to give an 1--5 rating to the statement ``Using \prodigy\ would make it easier for me to do process modeling tasks.'' The \href{https://github.com/Nouronihar/BPM24_LLM4PM/blob/main/evaluation_interview.pdf}{complete questionnaire} is provided in our repository~\cite{repository}.

\subsection{Dissemination and Communication of the Results}
\label{ssec:dissemination}

\vspace{-2mm}

\noindent
To conclude our DSR process, we formalized our learnings and made them accessible to interested parties. We documented our observations, analyzed our findings, identified lessons learned, stated limitations, and recommended directions for future work. We shared our learnings within Hilti Group and the wider BPM community in academia and practice---some companies reached out to us and expressed their interest about the development process of \prodigy.

\section{Key Findings and Lessons Learned}
\label{sec:findings}

\vspace{-2mm}

\noindent
We first summarise the major results of our user studies, and then outline our proposed ``operating model'' for our developed LLM-based chatbot, \prodigy.

\vspace{0mm}
{\setstretch{0.8}\textbox{{\small \textbf{Confidentiality Statement:} To protect the privacy of the participants to our user studies, we cannot reveal the full transcript of their interviews. However, we are able to answer questions about their generic viewpoint on some specific issues.}}}

\subsection{Preliminary Interviews: what do Hilti process modellers want?}
\label{ssec:preliminary}

\vspace{-2mm}

\noindent
These open interviews lasted for 60 minutes, and the results shed light on the pain-points and desiderata of our participants. We found that, before modelling a process, 60\% spend between 15--60 minutes \textit{searching} for documentation; and also 60\% spend between 5--60 minutes to \textit{review} such documentation. As a matter of fact, 90\% state that it is ``extremely important'' that an LLM-based chatbot has access to Hilti's documentation; however, we also found that, on a 1--10 rating (low to high) scale, the average usefullness of current Hilti's documentation is 6.7---indicating helpfulness, but with huge margins for improvement. Nonetheless, with respect to AI-related concerns, some stated that ``humans may misinterpret the AI's outputs'' or ``AI may negatively impact collaboration with colleagues'' or even about accountability (``the mindset that [the machine] does everything and we no longer have to worry about it is dangerous'').

\begin{cooltextbox}
\textbf{\textsc{T}{AKEAWAYS}.} After analysing all our responses, we identified two design objectives which we used as basis to develop our LLM-based chatbot, \prodigy.
\begin{itemize}
    \item The chatbot should \textit{support process modellers} in creating BPMN models. In doing so, the chatbot should hint at the larger picture, i.e., emphasize and guide in purpose, usage, and value creation of the resulting models.
    
    \item The chatbot should be able to access and utilize existing documentation, and hence \textit{be aware of organizational specifics}. Such knowledge should drive the formulation of the output, which will be tailored to the organization.
\end{itemize}
The name \prodigy\ stands for ``PROcess moDellIng Guidance for You''.
\end{cooltextbox}

\vspace{-2mm}

\subsection{Evaluation: what do Hilti process modellers say about \prodigy?}
\label{ssec:evaluation}

\vspace{-2mm}

\begin{figure}[!htbp]
    \vspace{-0.6cm}
    \centering
    \frame{\includegraphics[width=0.95\columnwidth]{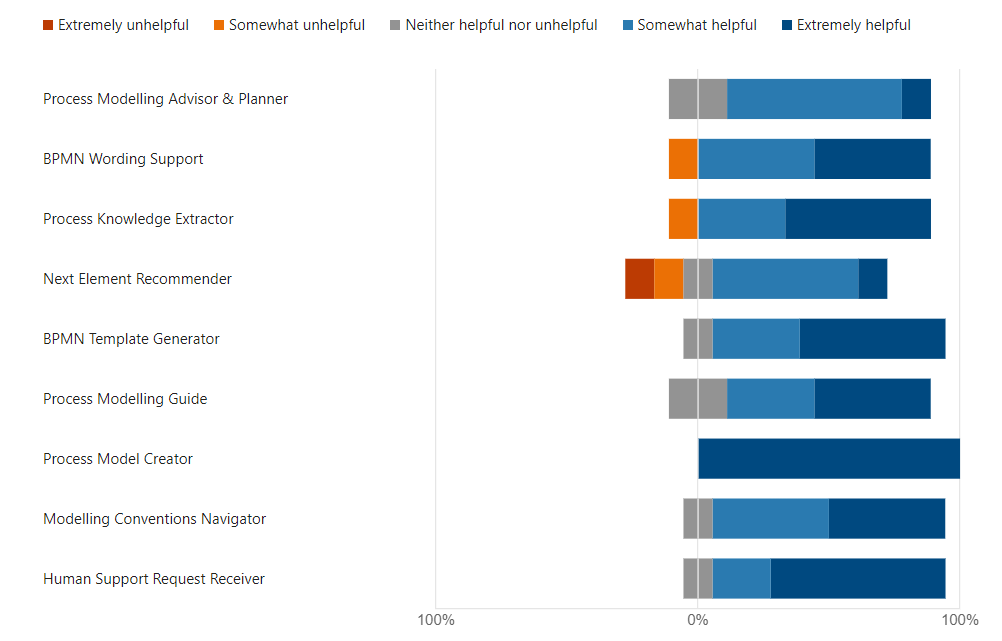}}
    \vspace{-0.3cm}
    \caption{\textbf{Helpfulness of \scprodigy's functionalities.} According to our participants, most of our implemented features are helpful---especially for \textit{creating process models} and \textit{supporting human requests}.}
    \label{fig:helpfulness}
    \vspace{-8mm}
\end{figure}

\noindent
After letting our process modellers use \prodigy, we collected their feedback via 90-minutes long semi-structured interviews; one participant to the preliminary interviews did not provide feedback since they were not available, so we obtained responses from nine employees. \textbf{The general opinion was positive}. Five of our participants asserted that they would use \prodigy\ on a daily or weekly basis (i.e., whenever they have to carry out process-modeling duties). Moreover, six participants asserted that \prodigy\ \textit{would speed-up} their tasks (three remained neutral), and eight believe that \prodigy\ \textit{makes their tasks easier} (one remained neutral). Finally, we report in Fig.~\ref{fig:helpfulness} the participants' perception on the functionalities we integrated in \prodigy, showing great appreciation. 

\subsection{Operating model: how should \prodigy\ be used in practice?}
\label{ssec:model}

\vspace{-2mm}

\begin{figure}[!htbp]
    \vspace{-0.6cm}
    \centering
    \includegraphics[width=1\columnwidth]{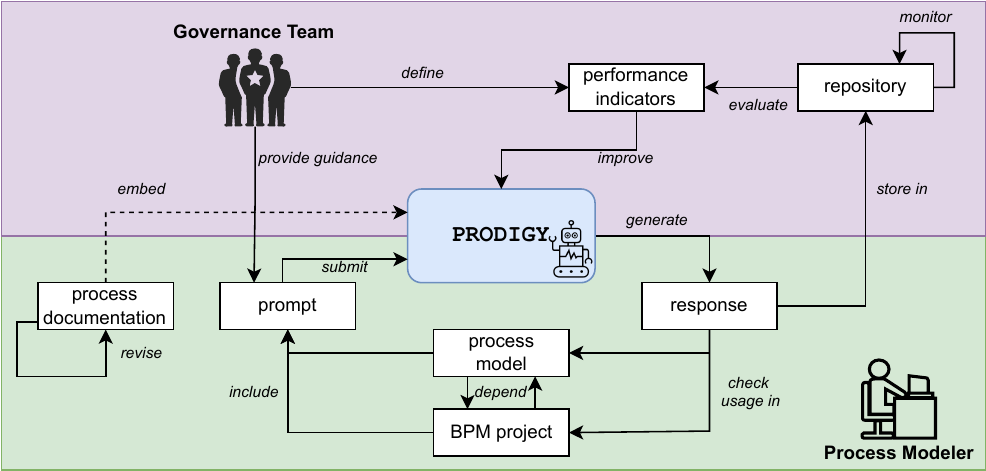}
    \vspace{-0.7cm}
    \caption{\textbf{Operating model of \scprodigy} --
    \textmd{\scriptsize We visualize the interactions between the \textit{Governance Team} (e.g., developers and managers) of a given organization with the \textit{Process Modeller} (i.e., the end-users of \scprodigy) that ensure a smooth operation of \scprodigy\ for real-world deployments.}}
    \label{fig:model}
    \vspace{-7mm}
\end{figure}

\noindent
During our implementation, we devised an \textit{operating model} that describes how \prodigy\ should be leveraged by real organizations; we have further refined our model (shown in Fig.~\ref{fig:model}) after receiving the feedback by our interviewees.

At a high-level, our model emphasizes continuous improvement through regular evaluations, feedback, and updates. To this end, our model delineates the interaction between a \textit{Governance Team} (i.e., the set of employees within a company that oversee the development and maintenance of \prodigy) and a \textit{Process Modeller} (i.e., the end-users of \prodigy). These two actors work collaboratively to ensure that the system performs well over time. 
For instance, the Process Modeller should be familiar with existing documentation and with the specific process, and scrutinize the response of \prodigy\ accordingly; they should also be willing to provide feedback (collected in a dedicated repository) and receive guidance from the Governance Team---who must, in turn, define clear performance indicators for \prodigy\ and periodically review the performance of prodigy (e.g., by analysing logs~\cite{stein2021towards_pc}) and apply updates if needed; as well as ensure that existing documentation is properly embedded in \prodigy\ (in a timely manner).

\section{Significance and Relevance in Research and Practice}
\label{sec:significance}

\vspace{-2mm}

\noindent
Besides our key findings we underscore three orthogonal aspects of our research.

\begin{figure}[!htbp]
    \vspace{-0.6cm}
    \centering
    \includegraphics[width=0.75\columnwidth]{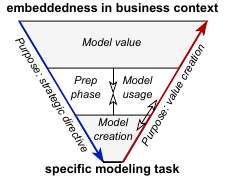}
    \vspace{-0.3cm}
    \caption{\textbf{Issues of process modellers.} Our framework has three levels with two-way transitions between each level---each having its own set of issues. \textit{Model value$\rightarrow$Model creation:} the organization must communicate a clear strategic directive for process model creation, balancing the cost/benefit of model creation and usage [issue: during the ``Prep phase'', process modellers find existing communication to lack clarity, leading to time waste and unproductive discussions among various stakeholders].  \textit{Model creation$\rightarrow$Model value:} While creating the model, process modelers should have a clear vision of ``who and how'' is going to use the model (which is what leads to the model becoming valuable) [issue: lack of clarity and/or poor documentation may lead to process models representing ``standalone exercises'' which do not bring any value to the company.]} 
    \label{fig:framework}
    \vspace{-7mm}
\end{figure}

\textbf{\underline{The perspective of process modellers in organizations.}}
We coalesce the responses---not pertaining to AI---of our preliminary interviews, and derive an \textit{original framework} representing dynamics of process modellers' issues at Hilti. This is instructive because, during our literature analysis, we found some works mentioning pitfalls of process modeling (e.g.,~\cite{rosemann2006potential}) but without accounting for context. Our framework (displayed in Fig.~\ref{fig:framework}, and described in the caption of Fig.~\ref{fig:framework}) attempts to rectify this shortcoming, providing guidance for future work.

\textbf{\underline{Evaluating LLM-/AI-based solutions.}} Upon further analysing the results of our evaluation interview, we have found that the reception of \prodigy\ by our process modellers was highly dependant on their expectations and overall attitude towards AI and IT innovation. Indeed, some participants had ``lower expectations'' and provided prompts that were ``more aligned'' to the expected input of \prodigy---and these participants rated \prodigy\ more positively. In contrast, participants who were expecting that \prodigy\ would ``do their work for them'' by issuing a single (and typically poorly phrased and/or unclear) prompt were more skeptical of \prodigy's helpfulness. These results underscore the importance of {\small \textit{(i)}}~accounting for each end-user's expectations while evaluating the performance of operational AI-based solutions; as well as {\small \textit{(ii)}}~educating end-users on the potential (and limitations) of AI-based tools.

\textbf{\underline{The role played by higher education.}}
With this paper, we (also) seek to bridge three domains: industrial practice, scholarly literature, and higher-education institutions~\cite{senkus2023bridging_pc}. This research has been predominantly carried by Clara Ziche for her MSc. thesis, during which she was working part-time at Hilti. The development of \prodigy\ was driven by following the guidelines of prior academic literature, and the resulting artifact was appreciated by Hilti as well as by other companies that witnessed its capabilities. On this note, we find it instructive to trace the timeline of this research by outlining the path followed by Clara Ziche to bring our findings to light. In Sept--Dec 2023, after attending the BPM'23 conference, Clara investigated the state of the art and designed the interviews for the requirement analysis. In Jan 2024, Clara carried out the interviews, and began familiarizing with current LLM technologies. In Feb 2024, Clara developed \prodigy\ and designed the questionnaire for its evaluation---which took place in March 2024. Disseminations occurred in Apr--May 2024.

{\setstretch{0.8}\textbox{{\small \textbf{Technical feasibility.} After having collected the input from experts, the development of \prodigy\ took only two weeks \textit{from a MSc. student in Information Systems}.}}}

\vspace{1mm}

\section{Discussion: Scope and Limitations}
\label{sec:discussion}

\vspace{-2mm}

\noindent
We showcased an exemplary application of an LLM-based chatbot that can assist process modellers in a large organization, Hilti. In doing so, we have carried out a twofold user study with 10 employees of Hilti, and developed an original artifact, \prodigy. Our research has a number of limitations. For instance, we do not claim that our findings can apply to other organizations---irrespective of their similarity to Hilti. Moreover, we cannot claim that even our own findings can apply to the entirety of Hilti: The participants of our user study are mostly based in Liechtenstein, and therefore cover the global headquarters perspective rather than regional and local perspectives. Furthermore, \prodigy\ uses GPT-3.5 Turbo (which is not privacy-compliant), and it relies on BPMN Sketch Miner: if such a tool is taken down, the output of \prodigy\ may lose its immediate usefulness. Finally, even our own participants have pointed out some shortcomings of \prodigy, such as a poor ``knowledge'' of Hilti's documentation. Such a result, however, was expected: the documents that \prodigy\ has access to (with RAG) are just a drop in the deluge of files and logs included in Hilti's databases (and we, as researchers, do not have complete access to such data). 
\section{Conclusions}
\label{sec:conclusions}

\vspace{-0.2cm}

\begin{figure}[!htbp]
    \vspace{-0.6cm}
    \centering
    \includegraphics[width=0.82\columnwidth]{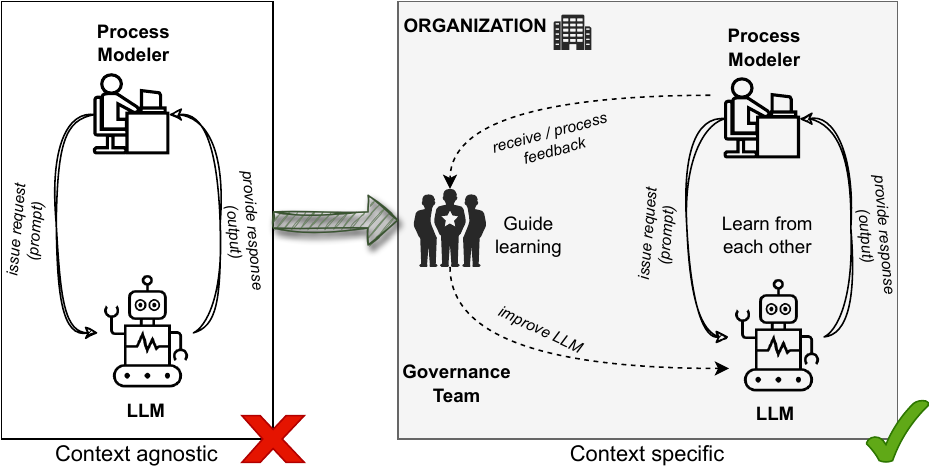}
    \vspace{-0.2cm}
    \caption{\textbf{Takeaway}. LLMs are hardly usable for process modeling in a context-agnostic setting [left]. Deployment of LLM in organizations for process modeling should follow a context-specific approach, in which the governance team ensures that LLM and end-users ``learn from each other'' [right].}
    \label{fig:takeaway}
    \vspace{-7mm}
\end{figure}

\noindent
We have presented the first case study showcasing how LLM can be used for process modeling in large enterprises---specifically, Hilti Group. We follow DSR guidelines and develop an original LLM-based chatbot, \prodigy, which we test with professional process modellers from Hilti. Our findings revealed that end-users appreciate the functionalities of \prodigy. However, concerns were raised about the poor alignment of \prodigy's output with Hilti's specifications. Such a shortcoming underscores the importance of integrating LLM-based solutions with the organization's documentation---which is a task outside the responsibilities of process modellers. Hence, deployment of similar solutions in real contexts should be done with the support of the organization's governance team: it is unrealistic to expect that ``off the shelf'' solutions work properly to drive the process modeling routines of complex and large organizations (see Fig.~\ref{fig:takeaway}).

\textbf{Acknowledgements.} We thank Hilti for enabling and funding this research; and the participants to our user study for their contributions and feedback.

\bibliographystyle{splncs04}

\end{document}